# A macro-modelling continuum approach with embedded discontinuities for the assessment of masonry arch bridges under earthquake loading

B. Pantò[1], S. Grosman[1], L. Macorini[1], B.A. Izzuddin[1]

[1] Department of Civil and Environmental Engineering, Imperial College London

South Kensington Campus, London SW7 2AZ, United Kingdom

## ABSTRACT

The paper presents an effective macro-modelling approach, utilising an anisotropic material model with embedded discontinuities, for masonry arches and bridges under cyclic loading, including dynamic actions induced by earthquakes. Realistic numerical simulations of masonry arch bridges under static and dynamic loading require accurate models representing the anionotropic nature of masonry and material nonlinearity due to opening and closure of tensile cracks and shear sliding along mortar joints. The proposed 3D modelling approach allows for masonry bond via simple calibration, and enables the representation of tensile cracking, crushing and shear damage in the brickwork. A two-scale representation is adopted, where 3D continuum elements at the structural scale are linked to embedded nonlinear interfaces representing the meso-structure of the material. The potential and accuracy of the proposed approach are shown in numerical examples and comparisons against physical experiments on masonry arches and bridges under static and dynamic loading.

**Notation**

Latin upper case letters

| | |
|---|---|
| **C** | Internal kinematics compatibility matrix |
| **D** | Damage matrix |
| $D_n$ | Damage index in the normal direction |
| $D_t$ | Damage index in the shear direction |
| $E_b$ | Young module of bricks |
| $E_m$ | Young module of mortar |
| $E_n$ | Homogenised normal module |
| $E_t$ | Homogenised shear module |
| $F_s$ | Yield surface in shear |
| $F_t$ | Yield surface in tensile |
| $F_c$ | Yield surface in compression |
| $G_b$ | Shear module of bricks |
| $G_m$ | Shear module of mortar |

$G_s$     Shear fracture energy

$G_t$     Tensile fracture energy

$G_c$     Compression fracture energy

**K**     Macroscopic tangent stiffness matrix

$\mathbf{K}_{int}$     Tangent stiffness matrix at the local level

**S**     Tangent stiffness matrix of a single Internal Layer

$\mathbf{S}_0$     Elastic stiffness matrix of a single Internal Layer

Latin lower case letters

$c$     Cohesion

**d**     Macroscopic strain vector

**e**     Strain vector at the local level

$g_s$     shear plastic potential

$f_t$     Tensile strength

$f_c$     Compression strength

$h_{bw}$     Bandwidth length

| | |
|---|---|
| $k_n$ | Normal stiffness of the mesoscale interfaces |
| $k_t$ | Shear stiffness of the mesoscale interfaces |
| $q$ | Hardening parameter of the damage-plasticity constitutive model |
| **s** | Stress vector at the local level |

Greek upper case letters

| | |
|---|---|
| $\Gamma_k$ | internal layer (IL) normal to the local *k* direction (k=x,y,z) |

Greek lower case letters

| | |
|---|---|
| **ε** | Internal Layer strain vector |
| **σ** | Internal Layer stress vector |
| **ε$_p$** | Internal Layer plastic strain vector |
| **$\tilde{σ}$** | Internal Layer effective stress vector |
| α$_{hk}$ | Parameter linking the macroscopic shear strains to the shear strains of the internal layers (hk=xy, xz, yz) |
| $\nu_b$ | Poisson's coefficient of bricks |
| $\nu_m$ | Poisson's coefficient of mortar |

$\phi$  friction coefficient

$\phi_g$  dilatancy coefficient

$\mu$  Model parameter of the damage-plasticity constitutive model

governing he closure of the tensile cracks

# 1 INTRODUCTION

Masonry arch bridges are old structures which still play a crucial role within modern railway and roadway networks. The vast majority of existing masonry bridges were built more than one hundred years ago mostly following empirical rules (McKibbins et al. 2006), and in many cases they date back to the medieval or Roman times representing important architectural heritage assets for numerous countries worldwide. Since their construction, they have been subjected to environmental actions causing progressive material degradation eventually leading to a reduction of the structural performance (Ural et al., 2008, Oliveira and Lourenço, 2004).

In general, masonry bridges show a very complex 3D response under traffic and extreme loading. It is governed by material nonlinearity in masonry, which in turn depends upon the properties of units and mortar joints and the bond pattern, and by the interaction among the different bridge components including arch, backfill, spandrel walls and piers in the case of multi-span bridges. Under dynamic loads, such as those inducted by earthquakes, the response is determined also by mechanical degradation with reduction of strength and stiffness of masonry due to opening and closure of tensile cracks and shear sliding along the mortar joints, by energy dissipation associated with the hysteretic behaviour of the backfill, and by the free-field conditions at the bridge boundaries.

Despite an increasing attention from the scientific community towards the behaviour of masonry arch bridges over the past three decades, many aspects concerning the cyclic and dynamic response of these complex structural systems are yet to be fully explored. Thus far, most of the research has been devoted to studying masonry bridges under static forces representing gravity and traffic loading. Laboratory tests were performed to investigate the load capacity of arch bridges (Sarhosis et al., 2016) evaluating the influence of backfill (Augusthus-Nelson et al., 2018; Callaway et al., 2012, Gilbert et al., 2007a; Melbourne et al, 1997), spandrel and wing walls (Royles and Hendry, 1991), and degradation phenomena associated with masonry bond, such as ring separation in multi-ring arches (Melbourne et al., 2007; Melbourne et al., 2004; Melbourne and Gilbert, 1995). On the other hand, very limited experimental studies focused on the response of masonry arches and vaults under horizontal cyclic loading (Gattesco et al., 2018) and dynamic actions (Giamundo et al., 2018), while the dynamic behaviour of masonry arch bridges under earthquake loading has yet to be investigated in physical tests.

On the numerical modelling front, masonry bridges are usually assessed employing 2D models based upon limit analysis concepts (Cavicchi and Gambarota, 2005; Gilbert et al., 2007b, Zampieri et al., 2016), the finite element method (Audenaert et al., 2008; Gago et al., 2011) or discrete element approaches (Sarhosis et al., 2019; Thavalingam et al., 2001). 2D descriptions generally entail a limited computational cost so they are suitable for practical assessment also of large bridges, but they cannot represent transverse behaviour induced by eccentric or lateral loading and the 3D interaction between the different bridge components. To overcome these limitations, 3D modelling strategies with different levels of detail and accuracy have been developed (Fanning et al., 2001; Milani and Lourenço; 2012, Conde et al., 2017, Zang et al., 2018; Tubaldi et al., 2018; Pulatsu et al, 2019; Caddemi et al., 2019; Zampieri et al. 2015), where the behaviour under earthquake loading has been studied mainly using FE continuum approaches with macroscale material descriptions for masonry (Pelà et al., 2009, Pelà et al., 2013; Conde et al., 2017). In these models, masonry is simulated by an equivalent homogeneous material and isotropic smeared-crack nonlinear constitutive laws, assuming plastic yield domains defined in terms of principal stresses, such as the total strain rotating crack model (TSRCM) implemented in DIANA FEA (2017). This model describes the tensile and compressive behaviour of the material employing a uniaxial stress-strain relationship and assuming that the crack directions rotate with the principal strain axes. Homogeneous finite element approaches have the advantage of requiring a limited computational effort, as the mesh size is independent from the actual masonry bond and the dimensions of the units. However, such macroscale material models bring important limitations, as they do not consider the anisotropy nature of brick/block masonry and do not allow for realistic damage accumulation under cyclic loading. Moreover, studies conducted on masonry buildings (Chisari et al., 2020), showed that macroscale isotropic models require complex calibration procedures to simulate the realistic brick/block-masonry response under cyclic loading conditions.

In this paper, the cyclic and dynamic behaviour of masonry arches and vaults is investigated employing a novel continuum macro-modelling strategy, which has been recently developed and employed to simulate the in-plane and out-of-plane response of masonry walls (Pantò et al., 2021). The proposed continuum model allows for masonry anisotropy by means of discrete embedded interfaces describing the response at the local level, while a continuum Cauchy representation is adopted at the macroscale. A simple but robust multi-scale approach is adopted to transfer information from the macroscale to the local level and *vice versa*. This modelling strategy enables a practical model calibration through direct use of mesoscale mechanical parameters, and it achieved a drastic reduction of the computational burden when applied to large structures, especially in comparison with detailed mesoscale approaches (Zang et al., 2018; Tubaldi et al., 2018). The ability of the proposed continuum macroscale description to predict the cyclic and dynamic response of masonry arches and bridges is evaluated considering 2D and 3D arch specimens, also interacting with backfill, and assuming as reference solutions the results from detailed mesoscale simulations. Parametric analyses are also performed to identify the most critical material parameters governing the hysteretic behaviour of the analysed systems.

## 2   MACROMODEL WITH EMBEDDED DISCONTINUITIES

The proposed macro-modelling strategy is based on a two-scale description of the masonry material. At the macro level, the masonry is simulated as an equivalent homogenized continuum material, discretised by means of a standard mesh of solid finite elements, while, at the local level, the meso-structure of the material is described by means of a uniform distribution of discontinuities, hereinafter referred to as *internal layers* (*ILs*). The Ils represent embedded interfaces, which allow spreading the plastic flow and damage (in real structures mostly concentrated at the mortar joints) uniformly into the material volume. As a result, at each Gauss integration point of an element domain, three ILs are considered ($\Gamma_x$, $\Gamma_y$, $\Gamma_z$), whose normal directions are along the main local material directions (*x*, *y*, *z* in Figure 1) corresponding to the orientations of the mortar joints within the brick/blockwork. The orientation of the *ILs*, with reference to a multi-ring barrel vault, is schematically depicted in Figure 1. More specifically, the local *z* axis corresponds to the direction of the circumferential mortar joints connecting adjacent rings, while the *x* and *y* axes are set along the bed and head joints of the masonry vault. This allows the macro-model to effectively simulate the 3D anisotropic nature of the material, also in presence of double-curvature or irregular geometries such as skew arch textures (Sarhosis et al., 2104).

According to the proposed modeling strategy, a Cauchy continuum strain field in the local reference system, $\boldsymbol{\varepsilon} = [\varepsilon_x \quad \varepsilon_y \quad \varepsilon_z \quad \gamma_{xy} \quad \gamma_{xz} \quad \gamma_{yz}]^T$, describes the deformations at the macro level. The macroscopic strains are transferred to the mesoscale level at each Gauss point of the domain (*P* in Figure 2) and used to compute the deformations of the internal layers $\mathbf{d}_k = [d_k \quad d_{kh} \quad d_{kh}]^T$ (*k=x, y, z* ; *h=x, y, z with h≠k*) composed of one normal component ($d_k$) and two shear components ($d_{kh}$) on the plane with normal *k*.

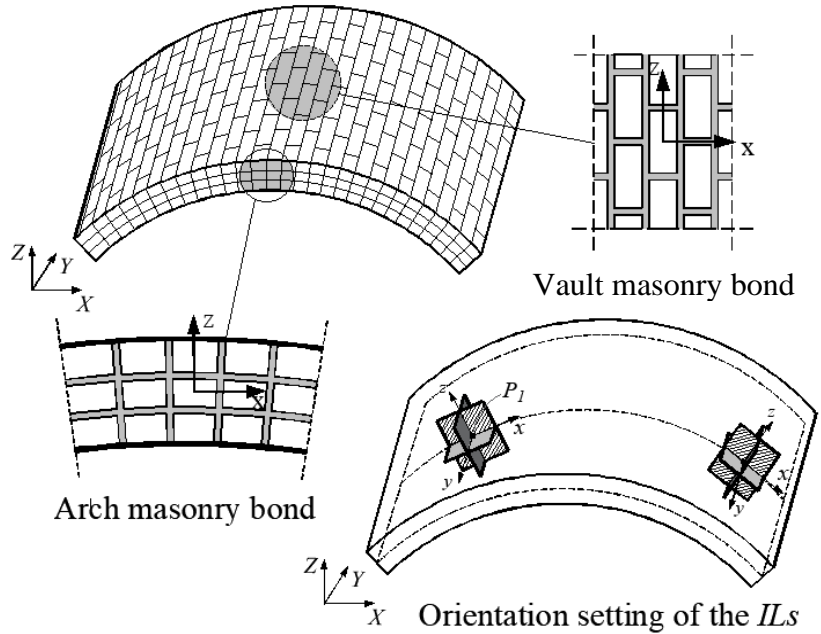

Figure 1. Orientation of the local layers according to the vault and arch masonry bond.

The dual local stresses of the ILs $\mathbf{s}_k = [s_k \quad s_{kh} \quad s_{kh}]^T$ are linked to the local strains by the incremental relationship $\delta \mathbf{s}_k = \mathbf{D}_k \delta \mathbf{d}_k$, where $\mathbf{D}_k$ is the tangent stiffness matrix associated with the plane $\Gamma_k$. The complete sets of local strains and stresses are ordered in the 9-components vectors $\mathbf{d}_{int}$ and $\mathbf{s}_{int}$:

$$\mathbf{d}_{int} = [d_x \quad d_y \quad d_z \quad d_{xy} \quad d_{xz} \quad d_{yz} \quad d_{yx} \quad d_{zx} \quad d_{zy}]^T \tag{1}$$

$$\mathbf{s}_{int} = [s_x \quad s_y \quad s_z \quad s_{xy} \quad s_{xz} \quad s_{yz} \quad s_{yx} \quad s_{zx} \quad s_{zy}]^T \tag{2}$$

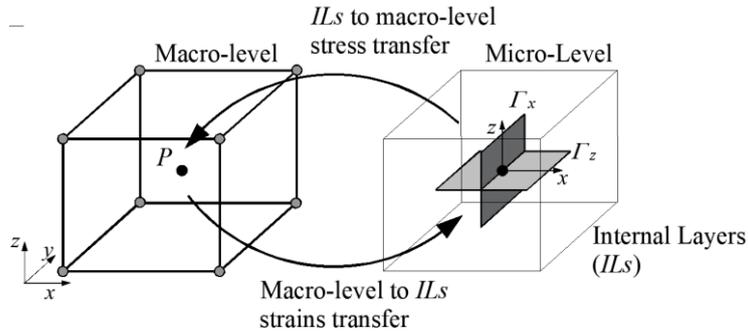

Figure 2. Schematic representation of the double-scale description adopted in the macro-model.

Simple kinematic compatibility relationships are utilised to link the local strains to the macroscopic strains, as expressed by:

$$\delta\mathbf{d}_{int} = \begin{bmatrix} 1 & & & & & & & & \\ & 1 & & & & & & & \\ & & 1 & & & & & & \\ & & & \alpha_{xy} & & & 1-\alpha_{xy} & & \\ & & & & \alpha_{xz} & & & 1-\alpha_{xz} & \\ & & & & & \alpha_{yz} & & & 1-\alpha_{yz} \end{bmatrix} \delta\boldsymbol{\varepsilon} \tag{3}$$

By imposing the Cauchy equilibrium conditions, namely the internal rotational equilibrium between the corresponding shear stress components ($\tau_{kh} = \tau_{hk}$), it is possible to evaluate the parameters $\alpha_i$ (i=1, 2, 3) corresponding to the increment of macroscopic strains ($\delta \boldsymbol{\varepsilon}$), and thus the relation linking the increment of the internal strains to the increment of the macroscopic strains (Pantò et al., 2021):

$$\delta \mathbf{d}_{int} = \begin{bmatrix} [\mathbf{I}_{3x3} \quad \mathbf{0}_{3x3}] \\ -\mathbf{A}^{-1}\mathbf{B} \\ [\mathbf{0}_{3x3} \quad \mathbf{I}_{3x3}] + \mathbf{A}^{-1}\mathbf{B} \end{bmatrix} \cdot \delta \boldsymbol{\varepsilon} = \mathbf{C}(\mathbf{D}_x, \mathbf{D}_y, \mathbf{D}_z) \cdot \delta \boldsymbol{\varepsilon} \qquad (4)$$

where:

$$\mathbf{A} = \begin{bmatrix} S_{x,22} + S_{y,22} & S_{x,23} & -S_{y,23} \\ S_{x,32} & S_{x,33} + S_{z,22} & S_{z,23} \\ -S_{y,32} & S_{z,32} & S_{y,33} + S_{z,33} \end{bmatrix} \qquad (5)$$

$$\mathbf{B} = \begin{bmatrix} S_{x,21} & -S_{y,21} & 0 & -S_{y,22} & 0 & 0 \\ S_{x,31} & 0 & -S_{z,21} & 0 & -S_{z,22} & -S_{z,23} \\ 0 & S_{y,31} & -S_{z,31} & SD_{y,32} & -S_{z,32} & -S_{z,33} \end{bmatrix} \qquad (6)$$

As can be observed from Eq. (4), the kinematics compatibility matrix (**C**), linking the macroscopic and local strains, depends on the tangent stiffness matrix of the ILs. As a result, the procedure is iterative. In the numerical application, a full Newton-Raphson scheme is adopted until the internal equilibrium is reached (Pantò et al., 2021). At each iteration, Eq. (4) is used to perform the elastic prediction phase, while the plastic correction is carried out by integrating independently the constitutive laws of the three ILs and evaluating the residual unbalance vector for the next iteration. More details can be found in (Pantò et al., 2021).

When local convergence is reached, the local macroscopic tangent stiffness matrix (**K**), which is required for determining the increment of nodal displacement within the FE solution procedure, is evaluated by imposing the virtual work principle resulting in:

$$\mathbf{K} = \mathbf{C}^T \cdot \mathbf{K}_{int} \cdot \mathbf{C} + \partial \mathbf{C}^T/\partial \boldsymbol{\varepsilon} \cdot \mathbf{s}_{int} \tag{7}$$

where $\mathbf{K}_{int}$ is the local tangent stiffness matrix containing the tangent stiffness matrices of the three internal layers. The second term in Eq. (7) is zero given that $\mathbf{s}_{int}$ satisfies the Cauchy equilibrium condition. Therefore, the stiffness matrix can be written as:

$$\mathbf{K} = \mathbf{C}^T \cdot \mathbf{S} \cdot \mathbf{C} \tag{8}$$

## 3  MACROMODEL CALIBRATION

In this work, the damage-plasticity material model developed by Minga et al. (2018) is adopted to describe the mechanical behaviour of the ILs. This underlying constitutive model considers three strain-stress components: $\mathbf{s} = [\sigma \quad \tau_1 \quad \tau_2]^T$ and $\mathbf{d} = [\varepsilon \quad \gamma_1 \quad \gamma_2]^T$ The concept of effective stress $\tilde{\mathbf{s}} = \mathbf{S}_0(\mathbf{d} - \mathbf{d}_p)$ representing the stress of a fictitious undamaged material is introduced. These stresses are evaluated by solving a linear-hardening elasto-plastic problem considering the elastic stiffness matrix $\mathbf{E}_0 = diag\{E_n \quad E_t \quad E_t\}$, with $E_n$ and $E_t$ the elastic normal and shear moduli of the material, and the plastic strains $\boldsymbol{\varepsilon}_p$. The nominal stress vector is obtained by multiplying the effective stress by the damage matrix $\mathbf{D} = diag\{D_n \quad D_t \quad D_t\}$, as given by:

$$\mathbf{s} = (\mathbf{I}_3 - \mathbf{D})\tilde{\mathbf{s}} = (\mathbf{I}_3 - \mathbf{D})\mathbf{E}_0(\mathbf{d} - \mathbf{d}_p) \tag{9}$$

The damage matrix contains the damage parameters in the normal ($D_n$) and shear ($D_t$) directions, ranging from 0 (no-damage) to 1 (complete damage). The damage evolution is governed by three ratios between the tensile, compressive and shear plastic works and the corresponding fracture energies ($G_t \quad G_c \quad G_s$). The damage in the normal direction assumes two different expressions in relation to the sign of the normal effective stress to allow the recovery of the normal stiffness in compression after the closure of tensile cracking (Minga et al., 2018).

Three plane yield surfaces define the elastic limits in shear ($F_s$) tension ($F_t$) and compression ($F_c$), as reported in the following:

$$F_s(\tilde{\boldsymbol{\sigma}}, q) = \sqrt{\tilde{\tau}_1^2 + \tilde{\tau}_2^2} + \tilde{\sigma} \tan(\phi) - c' \tag{10}$$

$$F_t(\tilde{\boldsymbol{\sigma}}, q) = \tilde{\sigma} - (f_t - q) \tag{11}$$

$$F_c(\tilde{\boldsymbol{\sigma}}) = -\tilde{\sigma} + f_c \tag{12}$$

A 2D representation of the yield domain in the space $\tilde{\sigma} - \tilde{\tau} = \sqrt{\tilde{\tau}_1^2 + \tilde{\tau}_2^2}$ is reported in Figure 3. In the expressions, $f_t$ and $f_c$ are the tensile and compressive material strengths, $\phi$ the friction angle and $q$ a linear hardening variable, ranging from 0 (initial value) to the limit value $q_{lim} = \frac{c}{tan}(\phi) - f_t$. Moreover, $c' = c$ if $q \leq q_{lim}$ and $c' = c + (q - q_{lim})tan(\phi)$ if $q > q_{lim}$. With the increase of $q$, the surface $F_t$ reduces until becoming a point when $q$ reaches the value $q_{lim}$. On the other hand, $F_s$ increases with the increase of $q$ if $q > q_{lim}$. Two distinct associated plastic flows are defined for $F_t$ and $F_c$, while a non-associated flow rule is assumed in shear employing a plastic potential ($g_s$), obtained from $F_s$ substituting $\phi$ with $\phi_g$, to take into account the effects of masonry dilatancy.

The complete formulation of the underlying constitutive model, including the hardening and the cyclic rules, can be found in (Minga et al., 2018).

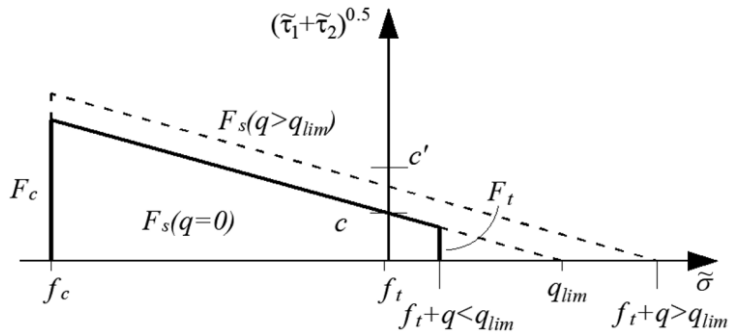

Figure 3. Yield surface used for the ILs.

The cyclic behaviour in tension is governed by the parameter ($\mu = \varepsilon_{p,f}/\varepsilon_f$), defined as the ratio between the residual normal strain ($\varepsilon_{p,f}$) when unloading form the point in which the full damage under pure tension ($D_n = 1$) is reached and the normal strain at that point ($\varepsilon_f$).

In this study, the model parameters of the constitutive law described above are evaluated following a simple calibration procedure, based on the mechanical properties of bricks and mortar joints or, alternatively, based on the corresponding mesoscale description of masonry. The elastic properties in the material local directions *x-y*, identifying the tangent plane of the masonry vault where a regular running bond is considered (Figure 4a), are evaluated following the homogenisation technique proposed in (Gambarotta and Lagomarsino, 1997). The expressions of the equivalent normal and shear module $E_{nx}, E_{tx}, E_{ny}, E_{ty}$ are reported in the following expressions

$$E_{nx} = \left[\frac{\mu_b}{E_b} + \frac{\mu_m}{E_m} - \frac{\mu_m \mu_b E_m E_b}{E_{nx}} \left(\frac{\nu_b}{E_b} - \frac{\nu_m}{E_m}\right)^2\right]^{-1} \tag{13}$$

$$E_{ny} = \mu_m E_m + \mu_b E_b \tag{14}$$

$$E_{tx} = E_{ty} = 2\left[\frac{2(1+\nu_b)\mu_b}{E_b} + \frac{2(1+\nu_m)\mu_m}{E_m}\right]^{-1} \tag{15}$$

where $\mu_m = \frac{h_m}{(h_m+h)}$ ; $\mu_b = \frac{h}{(h_m+h)}$

And: $E_b, G_b$ are the Young modulus and shear modulus of bricks; $E_m, G_m$ the Young modulus and shear modulus of mortar; $\nu_b$ and $\nu_m$ the Poisson's coefficients of bricks and mortar; $b, h$ the dimensions of the bricks and $h_m$ the average thickness of bed and head joints.

In the plane of the arch, a stack bond masonry, with no brick interlocking along the circumferential contact surfaces connecting the arch rings (Figure 4b), is considered as it is a common solution adopted for multi-ring arches. Consistently to this, the equivalent normal and shear moduli along the local *z* direction, are evaluated by combining in-series the stiffness of bricks and mortar, results in:

$$E_{nz} = \frac{E_m E_b (h_z + h_{mz})}{E_b h_{mz} + E_m h_z} \qquad (16)$$

$$E_{tz} = \frac{G_m G_b (h_z + h_{mz})}{G_b h_{mz} + G_m h_z} \qquad (17)$$

where, $h_z$ is the height of the bricks and $h_{mz}$ the thickness of the circumferential joints.

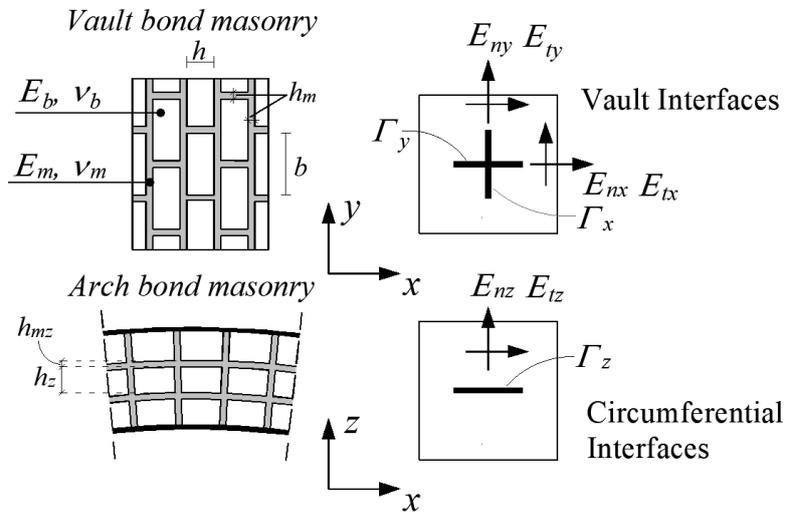

Figure 4. Macroscopic homogenization of the masonry periodic cell.

The nonlinear parameters along the local *x* and *z* directions, are considered coincident to the nonlinear parameters of the mortar joints. Namely, the tensile strength ($f_t$), cohesion ($c$), friction factor ($\phi$), and tensile and shear fracture energies normalised by the crack bandwidth $\left(\frac{G_t}{h_{bw}}, \frac{G_s}{h_{bw}}\right)$. The crack bandwidth, in each direction, is assumed as the minimum value between the mesh size and the dimension of the masonry unit in that direction.

The nonlinear parameters along the local y direction (transversal direction of the vault) are evaluated according to (Lourenco et al., 1997) taking into account the brick interlocking in this direction. The results are reported in Eqs (18-19) where $r_b = b/2h_m$ and the subscripts $j$ and $b$ indicate the parameters of bricks and mortar joints, respectively. Finally, the masonry strength in compression, in each direction, is assumed as a macroscopic parameter to be evaluated from experimental tests carried out on masonry prisms.

$$f_{ty} = min\{f_{tj} + c_j r_b \quad f_{tj} + f_{tb}/2\} \tag{18}$$

$$c_y = c_j \mu_m + c_b \mu_b \tag{18}$$

$$G_{ty} = \begin{cases} (G_{tj} + G_{sj} r_b)/(b + h_m) & if \quad f_{ty} = f_{tj} + c_j r_b \\ (G_{tj}\mu_m + G_{tb}\mu_b)/(b + h_m) & if \quad f_{ty} = f_{tj}\mu_m + f_{t,b}\mu_b \end{cases} \tag{19}$$

More accurate calibration procedures, based on 3D homogenisation techniques or numerical virtual tests, can be considered to improve the accuracy of the results. However, this task is beyond the scope of the present work which aims at investigating the accuracy and robustness of the proposed modelling strategy employing a simple calibration of the model parameters.

## 4 VERIFICATION, VALIDATION AND APPLICATION STUDIES

The proposed macro-modelling strategy has been implemented in ADAPTIC (Izzuddin 1991), an advance FE code for nonlinear simulations of structures, and used to investigate the response of bare masonry arches, 3D vaults, and arches interacting with backfill under monotonic and cyclic loadings. The numerical results have been compared against detailed mesoscale simulations according the approach proposed in (Macorini and Izzuddin, 2011) and against experimental data reported in the literature.

## 4.1   2D masonry arch

The first numerical-experimental comparison considers the two-ring brick-masonry arch, *Arch-G,* tested under vertical loading by Melbourne et al. (2007). The arch has 3 m span, 215 mm thickness and 455 mm width. It is made of class A engineering bricks 215×102.5×65 mm$^3$ arranged according to the stretcher bond with a continuous circumferential mortar joint connecting the two rings (Figure 5). Mortar joints are 10 mm thick and characterised by a volumetric cement:lime:sand ratio of 1:2:9. In the physical test, rigid reinforced concrete abutments were used to avoid support movements. Two initial vertical forces $F_0$ = 10 kN each were applied at quarter and three-quarter span and kept constant during the test. Subsequently, a vertical force (F) at quarter span was increased up to collapse under force control (Figure 5).

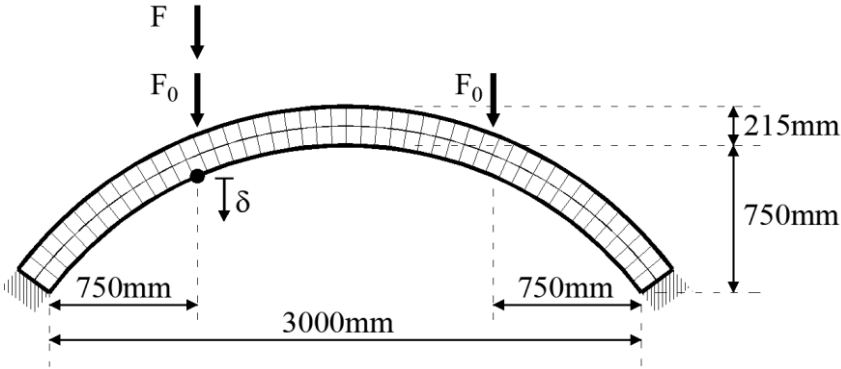

Figure 5.   Brick-masonry arch.

The developed mesoscale model is composed of 48 20-noded elastic elements simulating the bricks of each arch ring. Nonlinear interface elements (Macrorini and Izzuddin, 2011) are employed to describe the circumferential mortar joints connecting the two rings, and the bed mortar joints assumed as continuous along the radial directions. The mesoscale material parameters adopted in (Zhang et al. 2016) are used. More specifically, the solid elements are characterised by a Young modulus $E_b = 16000\ MPa$, a Poisson's ratio $\nu = 0.15$ and a specific self-weight of 22 kN/m$^3$. The mechanical parameters for the nonlinear interfaces are reported in Table 1, where, $k_n$ and $k_t$ are the normal and tangential stiffnesses, $f_t$, $f_c$ and $c$ are the tensile strength, the compressive strength and the cohesion, $\phi_t$ and $\phi_g$ are the friction and dilatancy angles.

Two different macroscale models with 1×24 and 2×36 20-noded solid elements have been developed to assess the influence of the mesh size on the arch response prediction. The macroscale mechanical parameters have been determined following the procedure described in Section 3. The homogenised linear parameters are reported in Table 2, where the nonlinear properties coincide with the nonlinear material characteristics of the mesoscale interfaces in Table 1. The effective lengths of 65 mm and 102.5 mm are used to normalise the fracture energies along the circumferential (*x*) and radial (*z*) directions, respectively.

Finally, the parameter $\mu$ governing the closure of the tensile cracks is assumed equal to 0.001, both in the mesoscale and macroscale description.

**Table 1**: Mechanical parameters of the meso scale model

| $k_n$ | $k_t$ | $f_t$ | $f_c$ | $c$ | $G_t$ | $G_s$ | $G_c$ | $tg\phi$ | $tg\phi_g$ |
|---|---|---|---|---|---|---|---|---|---|
| N/mm³ | N/mm³ | MPa | MPa | MPa | N/mm | N/mm | N/mm | - | - |
| 90.0 | 40.0 | 0.10 | 24.0 | 0.40 | 0.12 | 0.12 | 0.5 | 0.5 | 0.0 |

**Table 2**: Equivalent elastic propertes of the macromodel (MPa)

| Local direction z | | Local direction x | |
|---|---|---|---|
| $E_n$ | $E_t$ | $E_n$ | $E_t$ |
| 4818 | 4128 | 6956 | 3009 |

In the first simulation, the force *F* is monotonically increased reproducing the test protocol. The failure mechanisms obtained by the mesoscale model and the macroscale model with the finer mesh, with two solid elements along the thickness of the arch, are shown in Figure 6, where the von-Mises equivalent stress contours are also displayed. The two alternative numerical descriptions predict the same flexural mechanism with the activation of four radial cracks, as observed in the experimental test. The load-deflection curves showing the variation of the vertical displacement $\delta$ at quarter span underneath the load are displayed in Figure 7a. The macroscale curves obtained by using the two alternative meshes are in good agreement with the mesoscale response, confirming a negligible influence of the macroscale mesh characteristics. The macroscale models, however, overestimate the peak load of about 15% compared to the mesoscale and experimental results. A further study revealed that such discrepancy is associated with the contribution of the continuous circumferential mortar joint, which is spread along the entire thickness of the arch in the macroscale description. In fact, a mesoscale simulation for a single ring arch with the same geometrical and mechanical characteristics of the analysed two-ring masonry arch leads to a peak load very close to that predicted by the proposed macroscale model (Figure 7a). Nevertheless, for displacement values larger than the peak-load displacement (about 2mm), the two macroscale curves converge rapidly towards the mesoscale curve confirming an accurate prediction of the post-peak response.

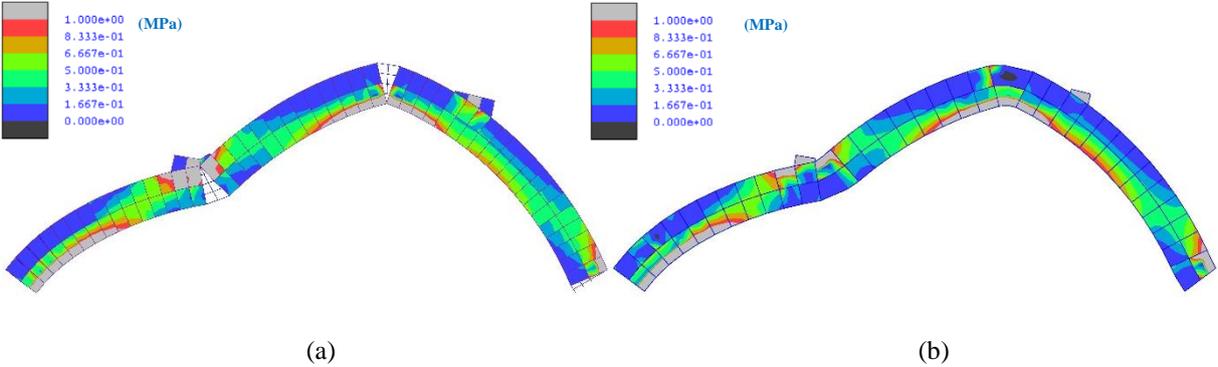

Figure 6. Ultimate deformed shapes and Von-Mises stress distribution of the mesoscale model (a) and macromodel (b).

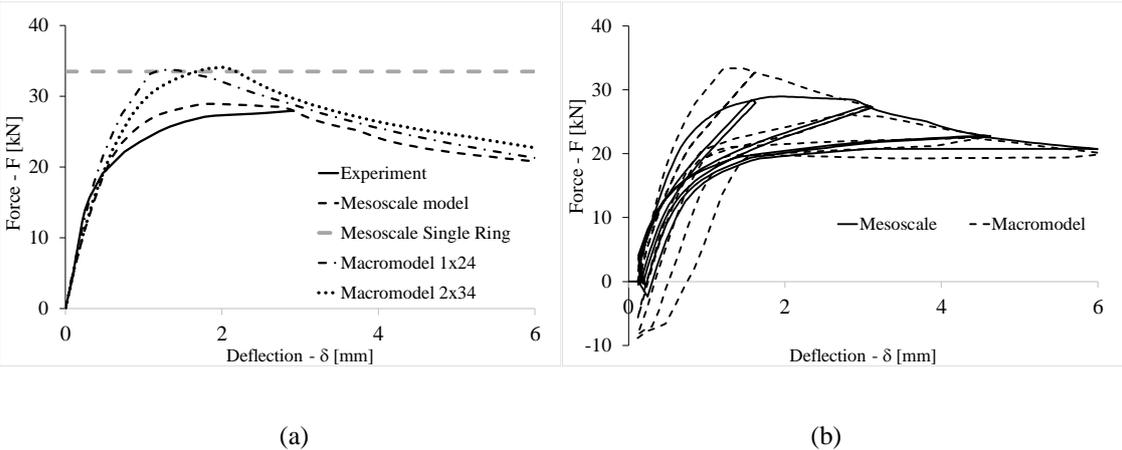

Figure 7. Force deflection curves of the arch, considering (a) monotonic and (b) cyclic loading paths.

In order to evaluate the ability of the proposed macroscale modelling strategy to predict the hysteretic behaviour of the arch, simulations under cyclic loading have been carried out using the 1×24 macroscale mesh and the mesoscale model. Four loading-unloading cycles with increasing maximum displacement (Figure 7b) have been applied. The two models predict a very similar hysteretic behaviour characterised by the opening and closure of tensile cracks. The main difference is observed in terms of residual deformations which are larger for the macroscale model.

## 4.2  Arch interacting with backfill

In this section, the single-span masonry bridge specimen *Bridge 3-1* tested by Melbourne and Gilbert (1995) is investigated using the proposed macroscale modelling strategy. The analysed structure comprises a two-ring arch, backfill and spandrel walls. The arch is 2880 mm wide and has the same span, rise and thickness of the arch shown in Figure 5 and considered in Section 4.1. The backfill extends horizontally 2460 mm from the two arch supports and is 300 mm deep at the arch crown. The *Bridge 3-1* specimen was subjected to a line load at the top surface of the backfill at the quarter span of the arch (Figure 8). The load was uniformly distributed along the width of the bridge inducing a cylindrical deformation mode up to failure. The spandrel walls were detached from the arch, but they provided lateral confinement to the backfill.

The specimen was analysed by Zhang et al. (2018) using a detailed mesoscale description explicitly allowing for the actual masonry bond. In this previous study, an efficient strip model representing a portion of the arch interacting with the backfill, which was fully restrained along the transverse direction, was adopted. A similar numerical model is used here, employing the same representation with 15-noded elasto-plastic tetrahedral elements for the backfill and a macroscale description for the arch, which corresponds to the coarser mesh with 1×24 20-noded solid elements introduced in Section 4.1. The adopted mesoscale and macroscopic masonry material properties are reported in Tables 1 and Table 2.

According to Zhang et al. (2018), an elasto-plastic material model with a modified Drucker-Prager yield criterion is employed for the backfill, assuming a Young's modulus $E_b$=500 MPa, a cohesion $c_b$=0.001 MPa, friction and dilatancy coefficients $tg\phi_b$=0.95 and $tg\psi_b$=0.45, and a specific self-weight of 22 kN/m$^3$. The interaction between the arch and the backfill is simulated by introducing nonlinear interfaces at the extrados of the arch with tensile strength $f_{fi}$=0.002 MPa, cohesion $c_i$=0.0029 MPa, friction coefficient $tg\phi_i$=0.6 and zero dilatancy. The backfill domain is fully restrained at the two bases and longitudinally restrained at the two lateral faces to represent rigid supports. In addition, the nodes on the two longitudinal faces are transversally restrained to simulate the lateral confinement provided by the spandrel walls.

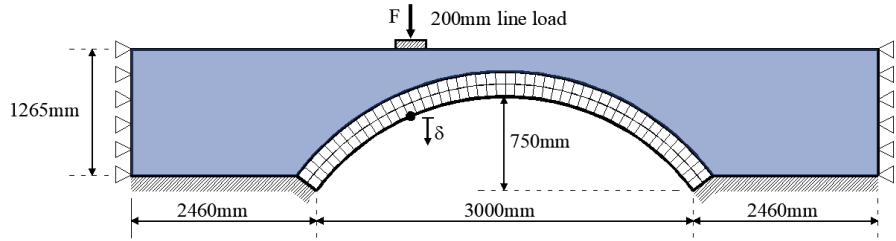

Figure 8. Geometry of the prototype and test layout.

The monotonic load-deflection curve obtained by the macroscale mesh is shown in Figure 9a, together with the experimental results and the mesoscale curve achieved adopting the same numerical description for the backfill. A very good agreement can be observed for displacement levels up to collapse. This improved accuracy compared to the results obtained for the bare masonry two-ring arch in Section 4.1, especially in term of peak-load prediction, corroborates the critical role played by the backfill and the arch-backfill interaction on the load-carrying capacity of masonry arch bridges.

As for the bare arch in Section 4.1, further nonlinear simulations have been conducted considering a cyclic loading-unloading history. The macroscale and mesoscale curves are shown in Figure 9b. Also in this case, the cyclic response of the macroscale model is in a good agreement with that predicted by the mesoscale model exhibiting a very similar hysteretic behaviour which is markedly different from that obtained for the bare arch, in section 4.1. This again emphasizes the importance function of the backfill and the importance of a realistic hysteretic model for the backfill to effectively simulate the dissipative capacity of the composite arch-backfill system under cyclic loading conditions.

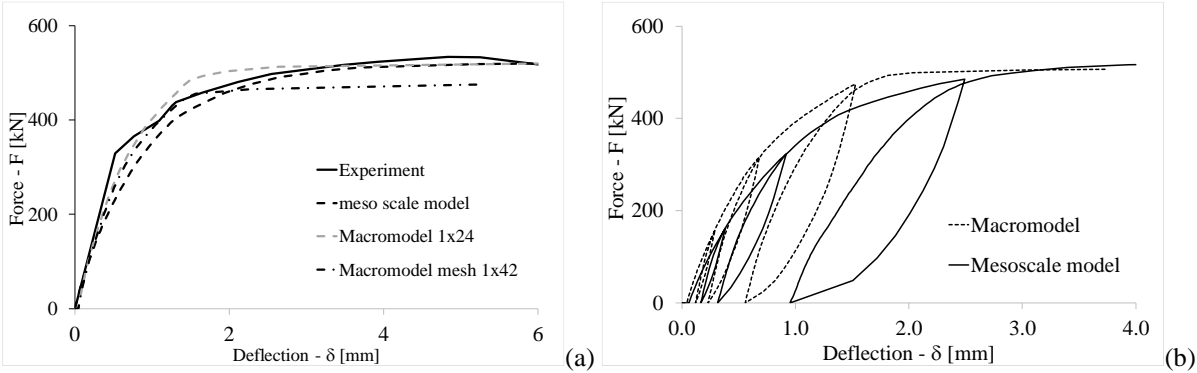

Figure 9. Force-deflection curves of the arch interacting with backfill: (a) monotonic and (b) cyclic loading.

Finally, Figure 10 displays the deformed shapes and the equivalent Von-Mises stresses for the mesoscale and macroscale models at collapse. In accordance with the experimental observations, both numerical descriptions predict a flexural mechanism with significant sliding between the arch and backfill on the side opposite to the loading area.

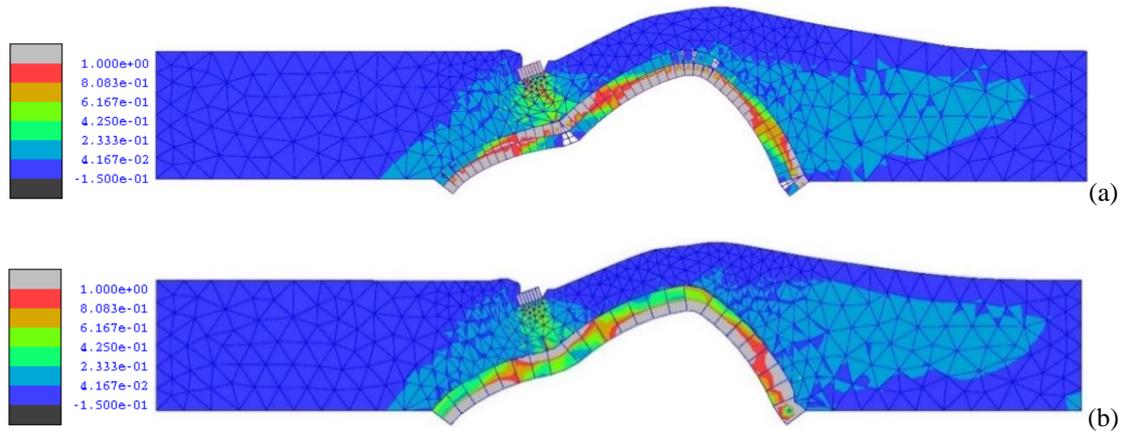

Figure 10. Deformed scape and Von-Mises stresses at collapse: (a) mesoscale model and (b) macroscale model.

## 4.3 Parametric analysis

To study the influence of the main model material parameters on the response under cyclic loading, parametric analyses have been carried out on both the bare two-ring masonry arch and the arch bridge specimen. The parameters considered in this numerical investigation have been selected amongst those that affect most the monotonic strength and stiffness of brick-masonry arches and bridges according to Tubaldi et al. (2020). In the case of the bare arch, the considered mesoscale material properties comprise the Young's module of the bricks ($E_b$), the normal stiffness ($k_n$), the tensile strength ($f_t$), the Mode-I fracture energy ($G_t$) and the residual strain factor ($\mu$) of the nonlinear interface elements representing mortar joints, whereas in the analysis of the arch with backfill also the influence of the backfill cohesion ($c_f$) has been investigated. The ranges of variation of the model material parameters are summarised in Table 3. Nonlinear simulations under displacement control have been carried out changing one parameter at a time. It has been found that a variation of $E_b$ and $\mu$ has only a marginal influence on the response predictions, but $k_n, f_t, G_t$ and $c_f$ are more influential.

**Table 3**: Variation of the model material parameters.

| $E_b$ (MPa) | | $k_n$ (kN/mm³) | | $f_t$ (MPa) | | $G_t$ (N/mm) | | $\mu$ (-) | | $c_f$ (MPa) | |
|---|---|---|---|---|---|---|---|---|---|---|---|
| $v_1$ | $v_2$ | $v_1$ | $v_2$ | $v_1$ | $v_2$ | $v_1$ | $v_2$ | $v_1$ | $v_2$ | $v_1$ | $v_2$ |
| 8000 | 40000 | 90 | 200 | 0.03 | 0.15 | 0.03 | 0.25 | 1E-4 | 0.1 | 0.5E-3 | 0.01 |

Figures 11 and 12 show mesoscale and macroscale load-displacement curves for the bare arch, respectively. It can be observed that both meso- and macroscale peak-load predictions are largely influenced by the variation of the tensile strength.

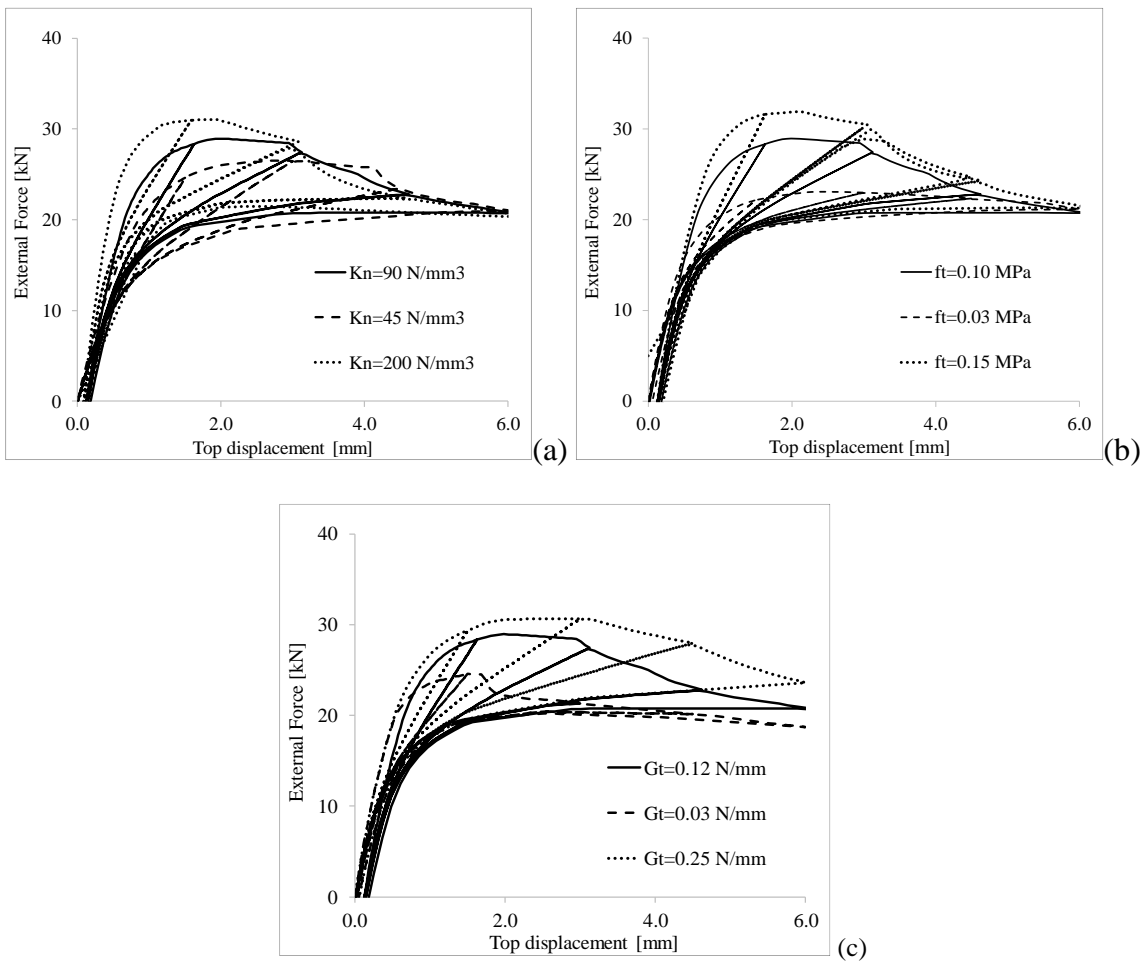

Figure 11. Mesoscale parametric simulations on the two-ring arch based on the variation of: (a) the normal stiffness; (b) tensile strength; (c) fracture energy of the nonlinear interface elements representing mortar joints.

Furthermore, both models lead to residual load capacity values which are less dependent from material parameters (apart from the case of large fracture energy) which indicates that residual strength depends mostly on the geometrical characteristics of the arch and the level of precompression induced by the gravity loads directly applied onto it. Comparing mesoscale and macroscale responses, it can be observed that the proposed macroscale modelling strategy, apart from overestimating the peak-load, predicts also an initial steeper post-peak softening branch and larger residual plastic deformations after unloading.

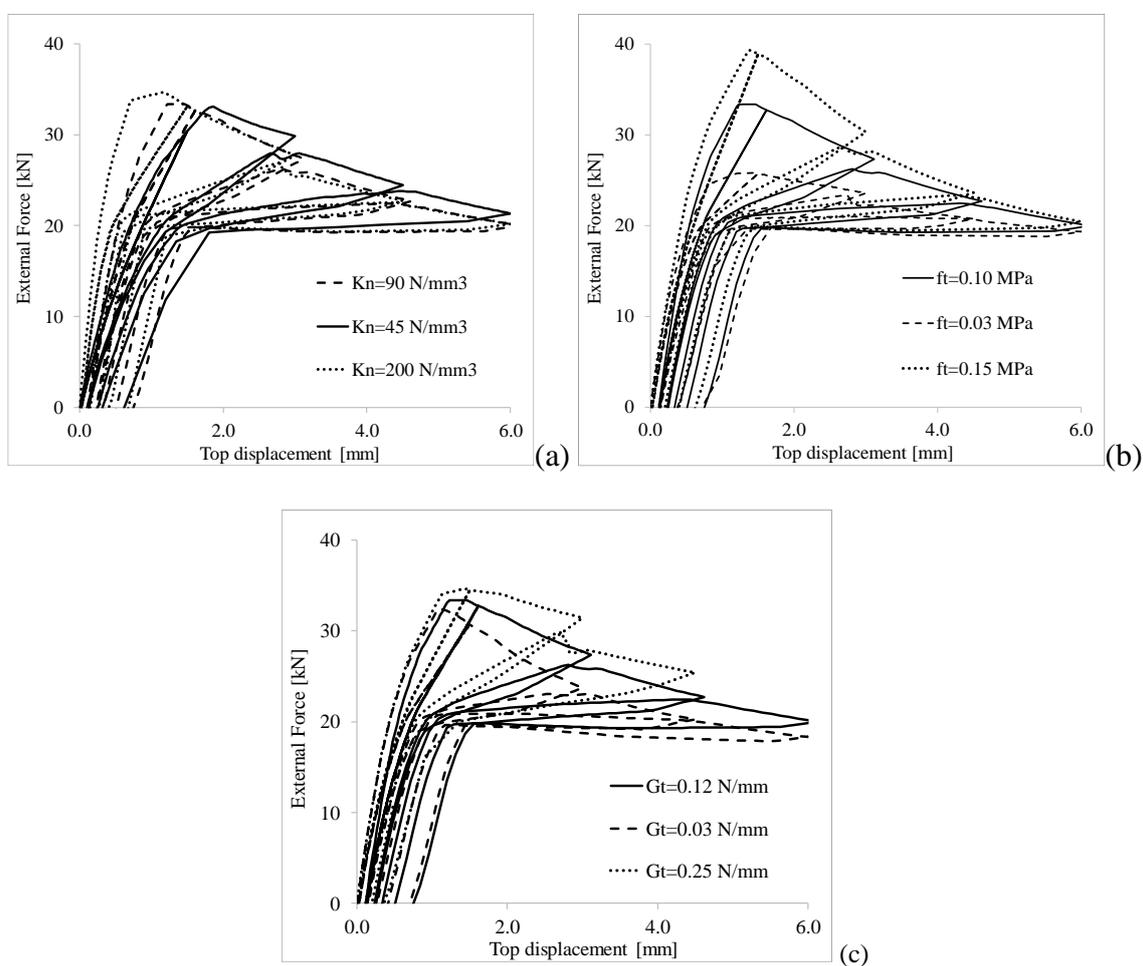

Figure 12. Macroscale parametric simulations on the two-ring arch based on the variation of the normal stiffness (a) tensile strength (b) and fracture energy (c) of the nonlinear interface elements representing mortar joints.

Figures 13 and 14 illustrate the main results obtained in the analysis of the arch with backfill. In this case, the backfill cohesion is the dominant parameter determining the peak load and the residual deformation upon unloading. Both mesoscale and macroscale curves are characterised by a mostly elasto-plastic behaviour with substantial residual deformations. This result is consistent with experimental observations on masonry arch bridge specimens subjected to vertical cyclic loading (Augusthus-Nelson et al., 2018) and confirm the major role played by the backfill which reduces the effects of potential variations of the masonry material characteristics.

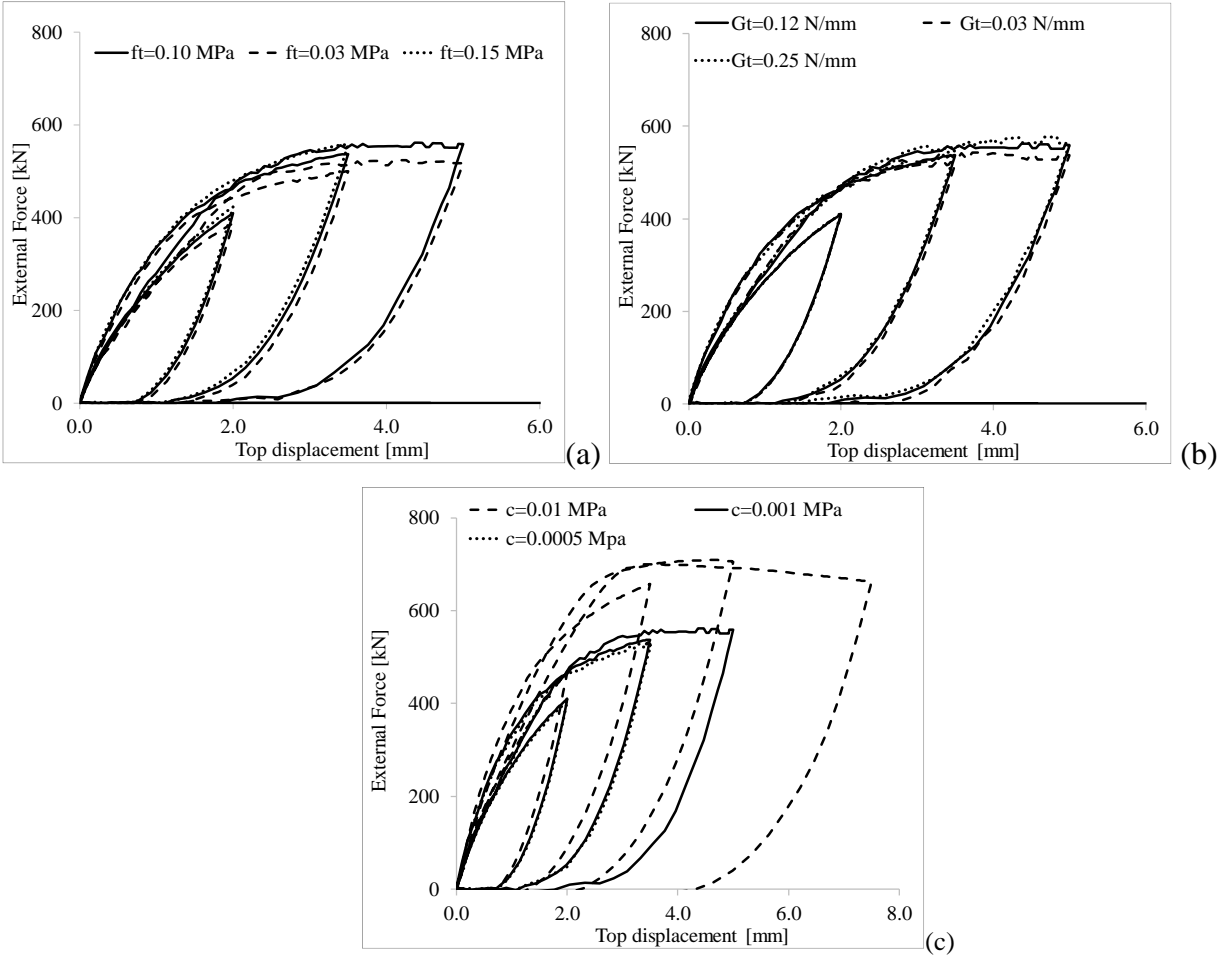

Figure 13. Mesoscale parametric simulations on arch with backfill based on the variation of tensile strength (a) fracture energy (b) and backfill cohesion (c).

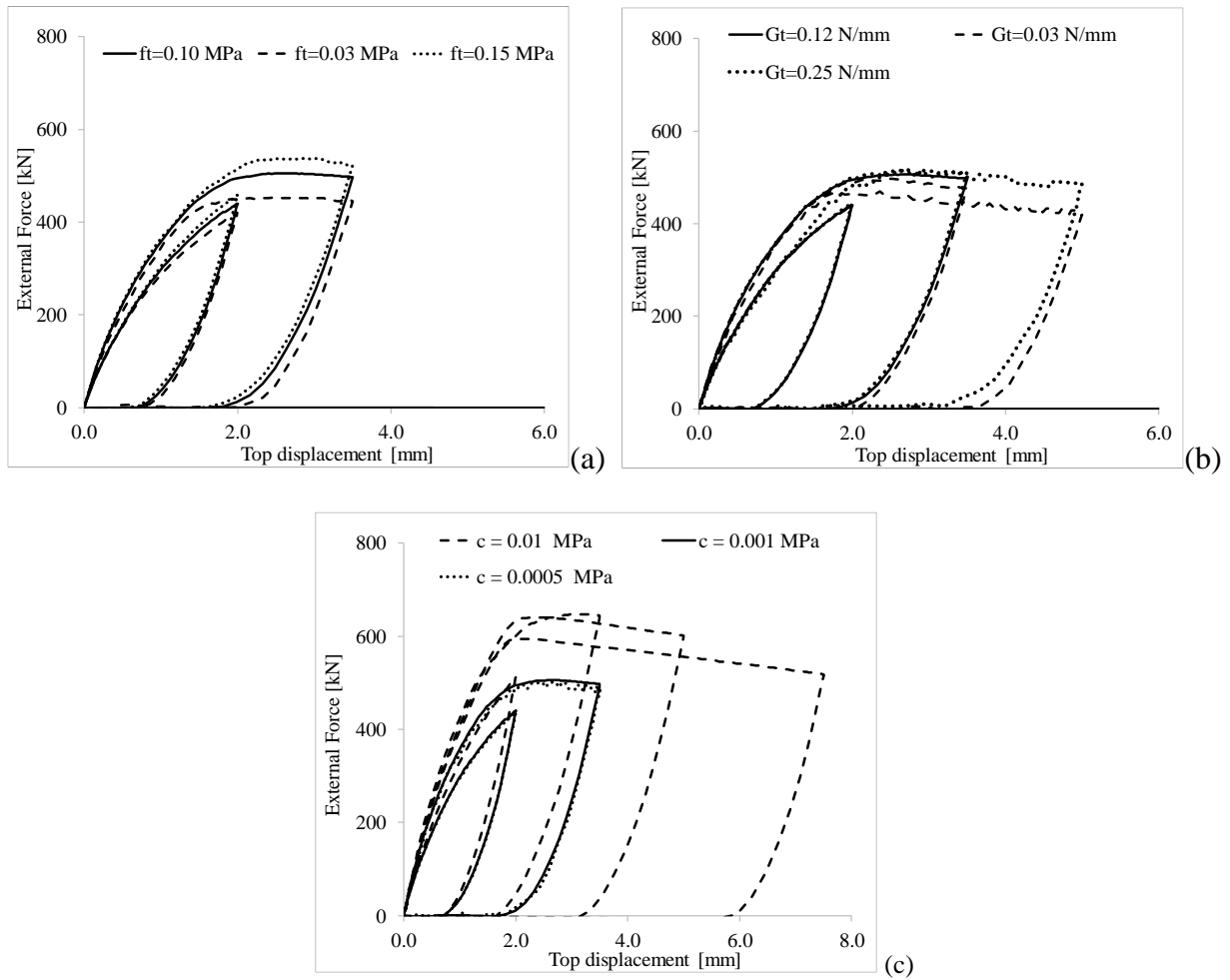

Figure 14. Macroscale parametric simulations on arch with backfill based on the variation of tensile strength (a) fracture energy (b) and backfill cohesion (c).

### 4.4 3D Masonry vault

The accuracy and potential of the proposed macroscale modelling strategy has been assessed investigating also the response of a 3D masonry vault under static as well as dynamic (earthquake) loading. As for the previous cases, the macroscale predictions are compared against detailed mesoscale simulations which provide baseline results for model validation. The analysed masonry vault is 3 m wide and has the same span, rise and thickness of the arch shown in Figure 5.

In the mesoscale model, the actual masonry bond along the transversal direction is also taken into account by representing each half brick with a 20-noded elastic element and alternating mortar interfaces with brick interfaces simulating the potential development of cracks within the bricks. The parameters reported in Table 1 are used for the mortar nonlinear interfaces, whereas the brick interfaces are characterised by high normal and shear stiffnesses $k_n = k_t = 10E5$ N/$mm^3$ and by the following strength parameters: $f_t = 2.0$ MPa, $G_t = 0.08$ N/mm, $c = 2.8$ MPa, $G_s = 0.5$ N/mm, $f_c = 24.5$ MPa, $G_c = 5.0$ N/mm, according to (Zhang et al., 2016). The macroscopic material properties indicated in Table 4 have been evaluated following the procedure described in Section 3.

In initial simulations, the vault has been subjected to static patch loads applied on an area of 300×400 $mm^2$, centred at three-quarter span (Figure 15). In a first test, the monotonic load is applied vertically ($F_v$) and the vault is fully restrained at the two bases, while the two lateral faces are either free or restrained in the vertical direction to simulate the effects of rigid spandrel walls. Subsequently, the vault is analysed under horizontal cyclic forces ($F_h$) with fixed supports at the bases and free at the two lateral faces.

**Table 4**: Mechanical properties adopted in the macromodel (MPa)

| Local Direction | $E_n$ [MPa] | $E_t$ [MPa] | $f_t$ [MPa] | $f_c$ [MPa] | $c$ | $G_t$ [N/mm] | $G_s$ [N/mm] | $G_c$ [N/mm] | $tg\phi$ [−] | $tg\phi_g$ [−] |
|---|---|---|---|---|---|---|---|---|---|---|
| x | 4818 | 4128 | 0.10 | 24.5 | 0.40 | 0.10 | 0.125 | 5.0 | 0.50 | 0.0 |
| y | 13980 | 4128 | 0.85 | 24.5 | 0.80 | 0.37 | 0.125 | 5.0 | 0.50 | 0.0 |
| z | 6956 | 3009 | 0.10 | 24.5 | - | 0.10 | - | 5.0 | - | - |

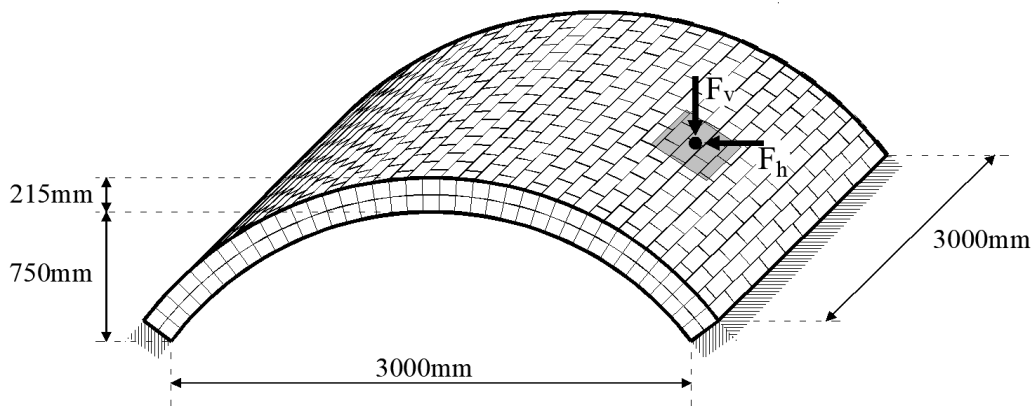

Figure 15. 3D spatial vault prototypes.

The numerical load-deflection curves obtained in the two numerical experiments are presented in Figure 16. More specifically, Figure 16a displays the results associated with the monotonic vertical load $F_v$, while Figure 16b shows the results for the horizontal cyclic load $F_h$.

In Figure 16a, MS-F and MS-R refer to the response curves from mesoscale models with free and restrained lateral faces, respectively, while MM-R and MM-F denote the corresponding response curves obtained with the macroscale models. The restrained models show a load capacity which is approximately double that predicted by the models with free ends and a reduced ductility. The proposed macroscale modelling strategy offers a very accurate prediction of the initial stiffness and the residual resistance for the two different support conditions, but it underestimates of about 17% the peak-load for the case with free end faces, and it overestimates the load capacity of about 7.5% for the case with vertically restrained end faces.

A very good agreement between mesoscale and macroscale results under horizontal cyclic loading can be observed in Figure 16b. This confirms that the proposed macroscale description enables a realistic representation of the cyclic hysteretic characteristics, including stiffness degradation and limited residual deformation upon unloading as predicted by the detailed mesoscale model.

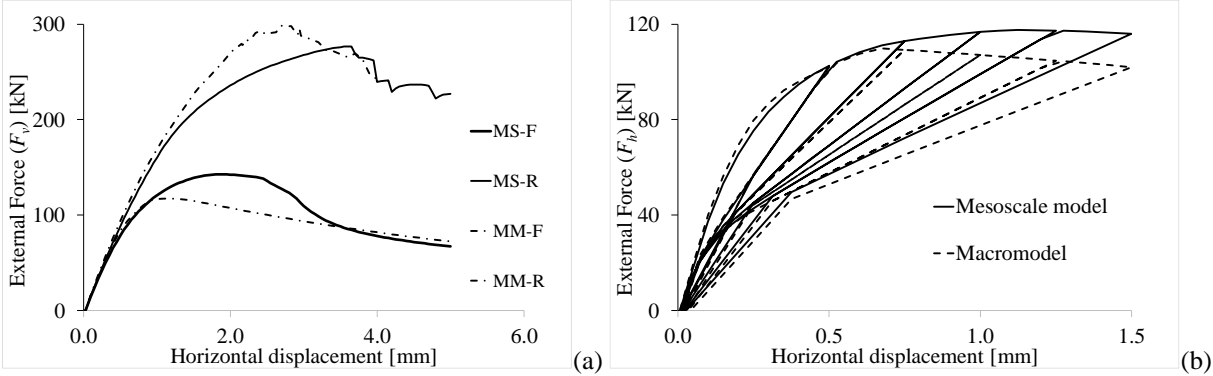

Figure 16. Load-deflection curves for the masonry vault under vertical (a) and horizontal cyclic loading (b)

Finally, the equivalent von-Mises stress distributions at the intrados and extrados of the vault with free end faces under vertical loading are shown in Figure 17. The stress pattern predicted by the macroscale model is consistent with that obtained by the mesoscale model, further confirming that the macroscale description leads to an accurate representation of 3D effects in masonry vaults.

In further numerical investigations, the vault has been subjected to prescribed acceleration histories at the two bases to represent the effects of earthquake loading. The N-S horizontal component of the Irpinia earthquake with PGA 0.32g (Figure 18) is simultaneously applied along the longitudinal and transversal directions of the vault. The earthquake record from Southern Italy Irpina earthquake (1980), Sturno station, is selected from the European Strong-Motion Database (ISESD), already processed by means of a linear baseline correction procedure and an eighth order elliptical bandpass filter with cut-off frequencies of 0.25 and 25.00Hz (Ambraseys et al. 2002; 2004).

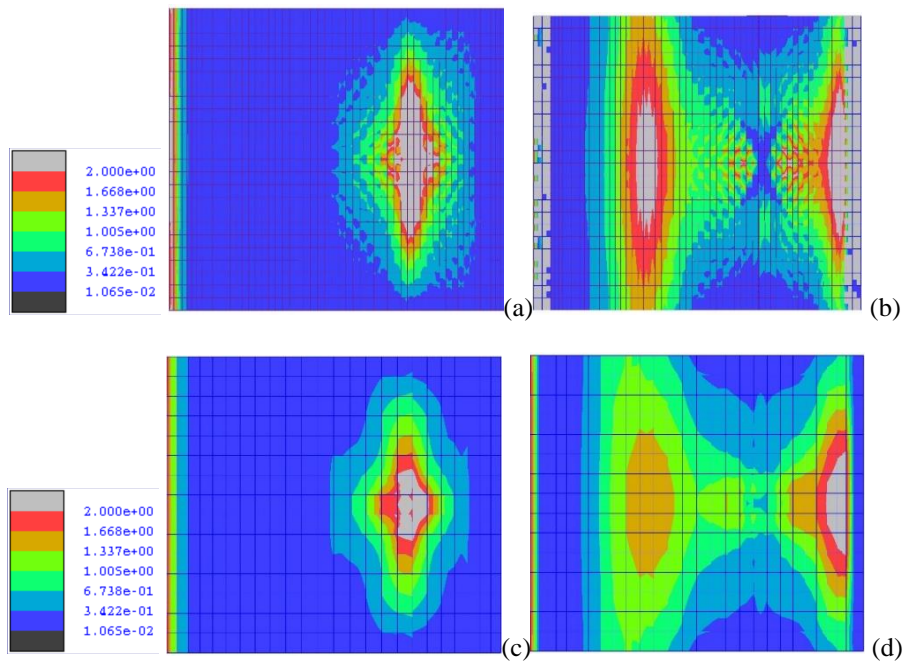

Figure 17. Equivalent von Mises stresses [MPa] at the peak-load for the vault with free ends under vertical loading: extrados (a) and intrados (b) of the mesoscale model and corresponding extrados (c) and intrados (d) of the macroscale description.

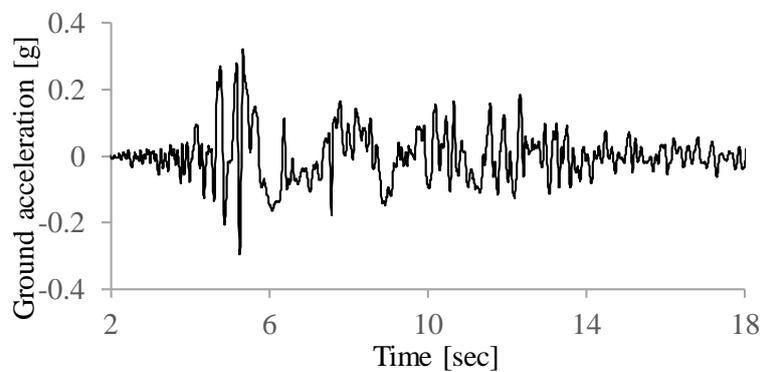

Figure 18. Acceleration ground motion history applied in the analyses.

A viscous damping proportional to the mass and associated with a damping coefficient $\xi=1\%$ for a reference period of 0.10 s is employed in the nonlinear simulations. This considers the effects of potential micro-cracks that are not taken into account by the adopted material description improving numerical stability. The original signal has been scaled until the failure of the vault which was reached by both models for PGA=2.56g (8 times the original PGA) at approximately 5.3 s. The failure mode predicted by the mesoscale model is characterised by shear sliding along the radial joints at the vault bases (Figure 19a). Extensive damage also develops along the vault span, as denoted by shear damage contours distribution at collapse shown in Figure 19b.

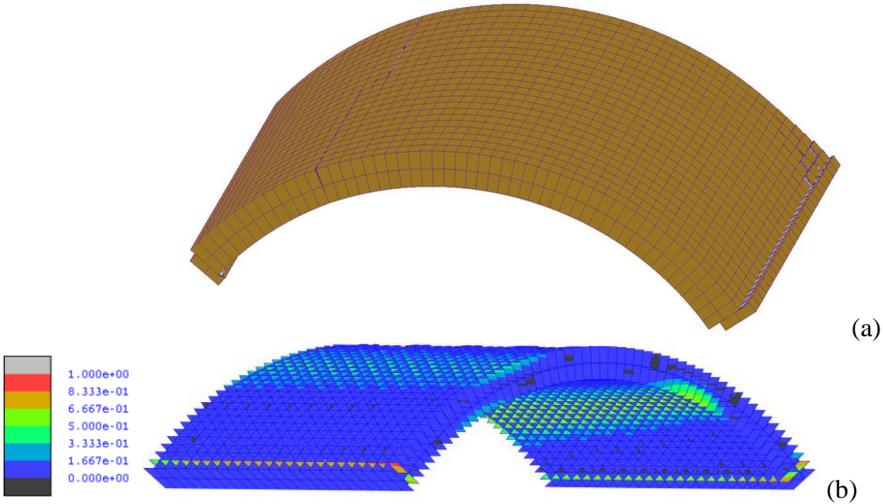

Figure 19. Mesoscale model: (a) sliding failure mechanism of the vaults; (b) shear damage distribution at the interfaces.

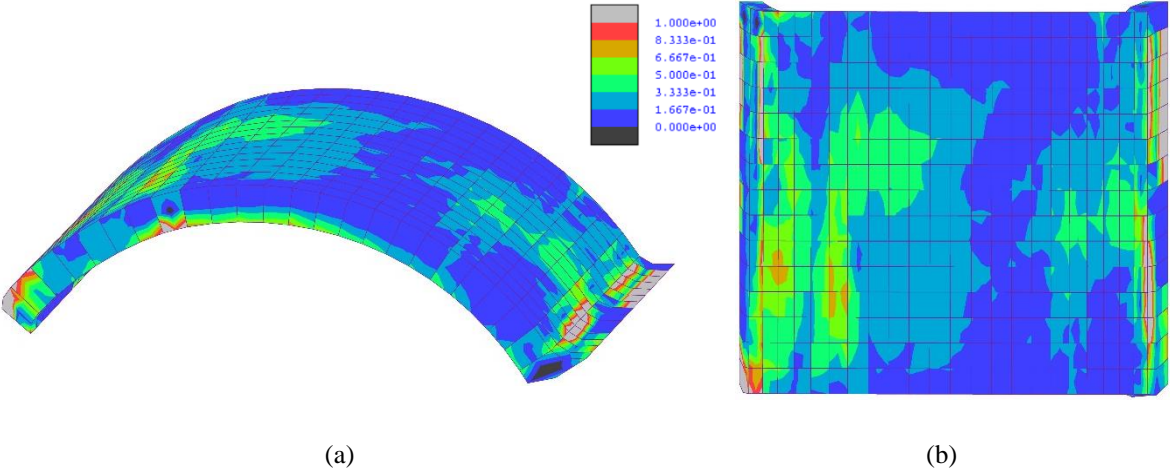

Figure 20. Deformed shape and Von-Mises stress distribution [MPa] of the macroscale model at collapse: (a) 3D and (b) top view.

A similar mechanism is predicted by the macroscale model as displayed in Figure 20, where significant shear deformations localised at the two supports.

Time-histories showing the variation in time of the relative longitudinal displacement centred at the crown of the vault, as predicted by the macroscale and mesoscale models for PGA=2.24g and PGA=2.56g (failure), are shown in Figures 21a and b, respectively. Also these results confirm the potential of the macroscale strategy and its ability in predicting the nonlinear dynamic response characteristics under earthquake loading, where the variation of peak displacements and frequencies are well simulated. The major differences from the mesoscale results can be noted from 6.0 s to 7.5 s, where the macroscale model shows higher high-frequency oscillations.

Finally, the hysteretic responses predicted by the two models for PGA=2.24g and PGA=2.56g are shown in Figure 22, where the variation of the base shear along the longitudinal direction is plotted against the longitudinal horizontal displacement centred at the crown. A general good agreement between the two descriptions can be observed for PGA=2.24g (Figures 17a,b). On the other hand, for PGA=2.56g, the macroscale model predicts wider hysteretic loops which may be due to the nature of the failure mode with a concentration of shear sliding at the two ends of the vaults which is spread within continuous elements in the macroscale model, whereas it is concentrated at the nonlinear interfaces representing the radial joints in the mesoscale description.

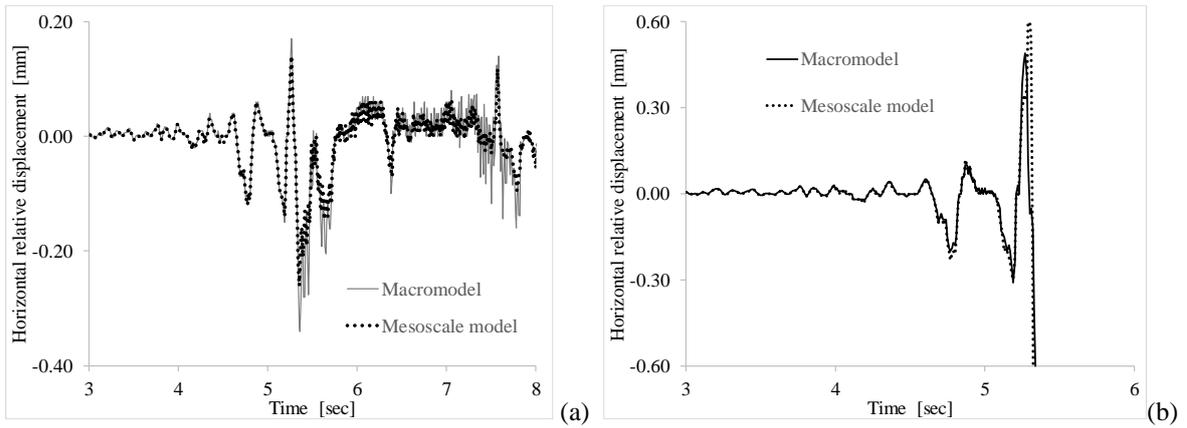

Figure 21. Displacement time-histories displacement corresponding to the ground motion with (a) PGA 2.24g and (b) PGA 2.56g.

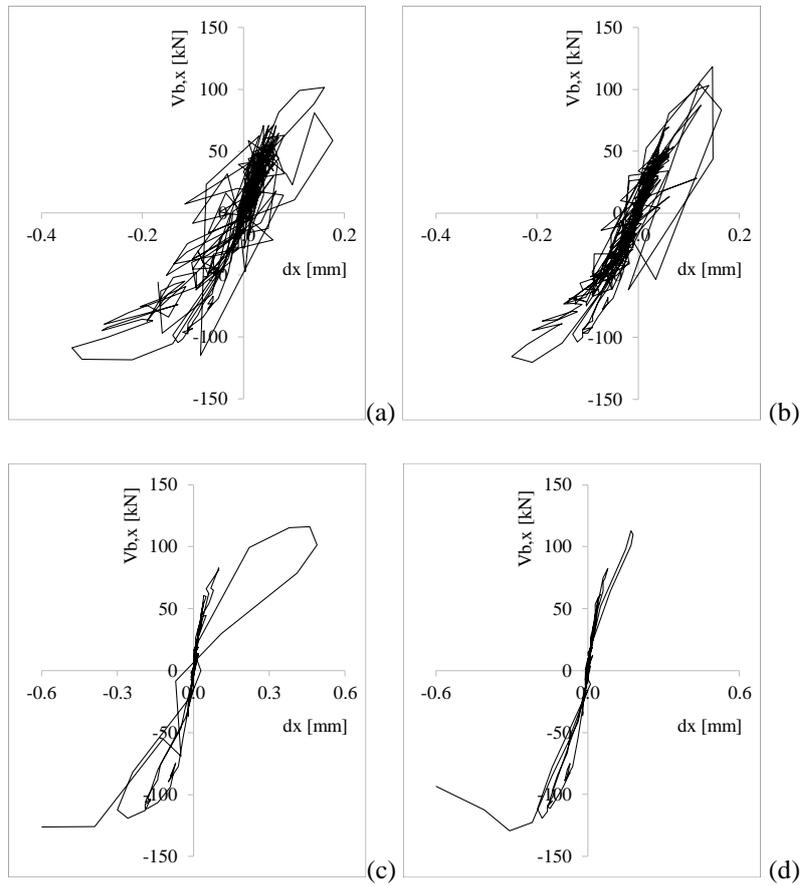

Figure 22. Hysteretic curves for macromodel (a) and mesoscale model (b) at PGA=2.24g; macromodel (c) and mesoscale model (d) at PGA=2.56g.

The computational times of the two description strategies have been registered to evaluate the efficiency of the macroscale model compared to the more detailed mesoscale approach. Regarding the time history analysis with PGA=2.24g, the macromodel showed an average speed-up value of approximately 83%. This result demonstrates the high potentiality of the macromodel to be adopted for the simulation of large structures.

In the next validation, a further vault specimen with the same span, thickness and mechanical properties of the vault analysed in the previous example, but with a different span-to-rise ratio of 2.5, is investigated applying the same accelerogram in both longitudinal and transversal directions. The displacement time-histories for the macroscale and mesoscale descriptions and different levels of PGA are reported in Figure 23. For PGA = 0.32g the two representations provide very similar responses. For higher values of PGA, 0.96g and 1.60g, the macromodel overestimates the displacement peaks, while providing a satisfactory prediction of the dynamic response of the vault in terms of frequencies of oscillation and predicting, coherently to the mesoscale model, the failure for a PGA of 1.60g. The hysteretic cycles of the two models for PGA=0.96g and 1.60g are reported in Figure 23. The hysteretic characteristics of the macromodel are consistent with those predicted by the mesoscale model, but the macromodel tends to overestimate the amplitude of the hysteretic loops, leading to an overestimation of the effective energy dissipation capacity of the structure. Finally, the failure mechanisms obtained by the two models are reported in Figure 25. In this case, a flexural failure mechanism is observed, which is due to the activation of four radial cracks. A very good agreement between the macroscale and mesoscale description can be observed.

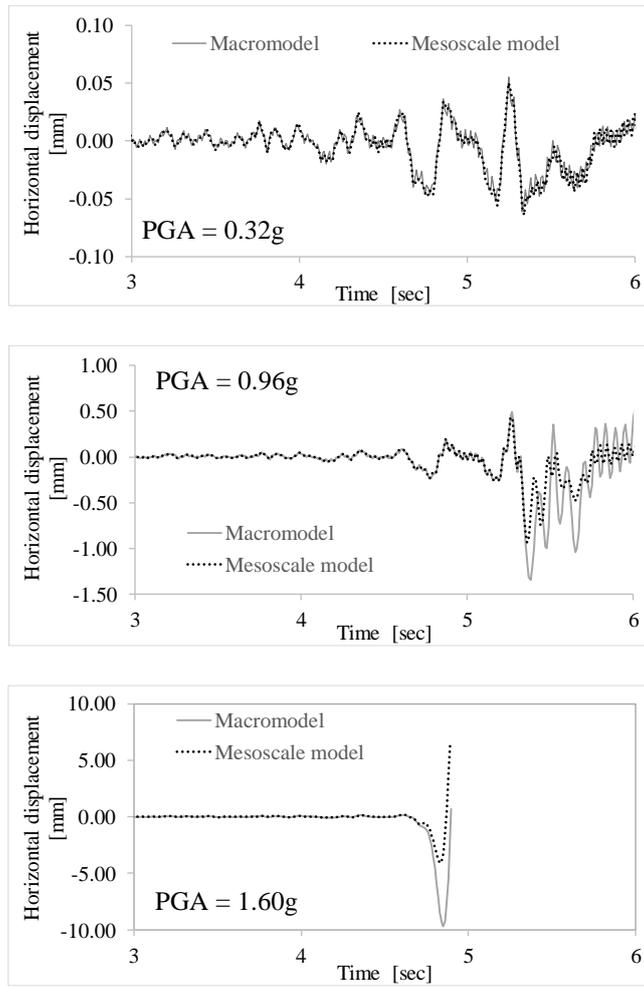

Figure 23. Displacement time-histories for different levels of PGA.

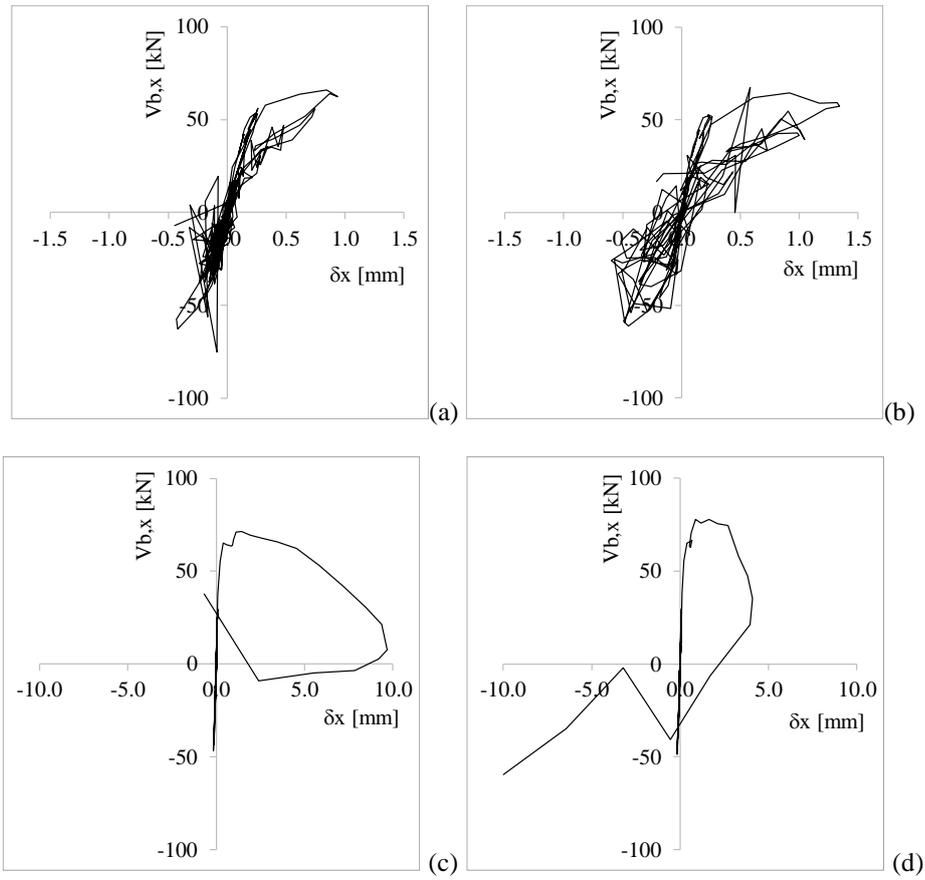

Figure 24. Hysteretic curevs for macromodel (a) and mesoscale model (b) at PGA = 0.96g; macromodel (c) and mesoscale model (d) at PGA = 1.60g.

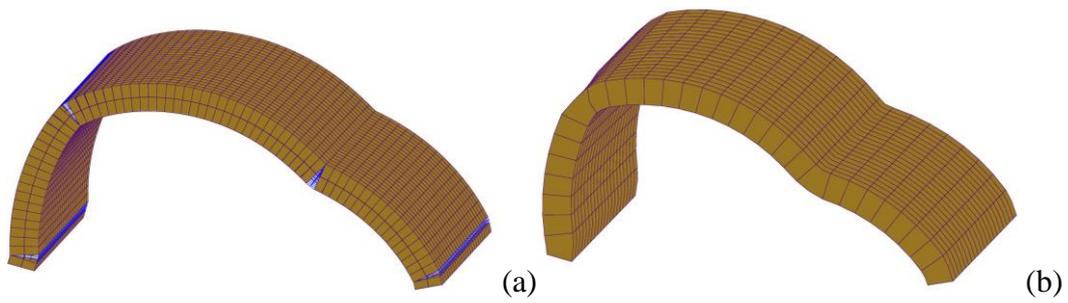

Figure 25. Failure mechanism predicted by the mesoscale model (a) and the micromodel (b).

# 5   CONCLUSIONS

The paper presents a novel 3D macroscale modelling approach for the simulation of masonry arches, vaults and arch bridges under static and dynamic loading. The proposed modelling strategy is based on a 2-level description of the masonry. At the macroscopic level, masonry is simulated as a homogeneous continuum Cauchy domain, whilst at the local level the mesoscale structure of masonry is represented by means of a discrete distribution of embedded interfaces (internal layers) simulating tensile, shear and compressive plastic deformations of mortar joints and potential cracks within bricks. A simple but robust calibration procedure, based upon the mesoscale mechanical parameters of masonry, is employed to evaluate the mechanical properties of the internal layers.

The proposed modelling strategy aims at providing an efficient but accurate numerical tool for the seismic assessment of large bridges with a reduced computational burden compared with detailed mesoscale descriptions. The model is validated with reference to 2D strip models of masonry arches, also interacting with a backfill layer, and spatial vault specimens subjected to static and dynamic loading conditions. The results of the proposed model are compared to those obtained by detailed mesoscale models and experimental data, demonstrating the ability of the developed modelling strategy to simulate the cyclic and dynamic response of masonry arches and arch bridges. Further investigations considering entire masonry bridges, including spandrel walls and piers, will be considered in future research.

**ACKNOWLEDGMENTS**

The first author gratefully acknowledges support from the Marie Skłodowska-Curie Individual fellowship under Grant Agreement 846061 (Project Title: Realistic Assessment of Historical Masonry Bridges under Extreme Environmental Actions, "RAMBEA", https://cordis.europa.eu/project/id/846061.